\documentclass[conference]{IEEEtran}
\IEEEoverridecommandlockouts
\usepackage{cite}
\usepackage{amsmath,amssymb,amsfonts}
\usepackage{algorithmic}
\usepackage{graphicx}
\usepackage{textcomp}
\usepackage{xcolor}
\def\BibTeX{{\rm B\kern-.05em{\sc i\kern-.025em b}\kern-.08em
    T\kern-.1667em\lower.7ex\hbox{E}\kern-.125emX}}

\usepackage{bm}

\usepackage[linesnumbered]{algorithm2e}

\usepackage{subcaption}

\newcounter{defcounter}
\begin{document}

\title{Optimal Management of Grid-Interactive Efficient Buildings via Safe Reinforcement Learning}


\author{Xiang Huo$^{\dagger}$, Boming Liu$^{\ddagger}$, Jin Dong$^{\ddagger}$, Jianming Lian$^{\ddagger}$, and Mingxi Liu$^{\mathsection}$
\thanks{*This manuscript has been authored by UT-Battelle, LLC, under contract DE-AC05-00OR22725 with the US Department of Energy (DOE). The US government retains and the publisher, by accepting the article for publication, acknowledges that the US government retains a nonexclusive, paid-up, irrevocable, worldwide license to publish or reproduce the published form of this manuscript, or allow others to do so, for US government purposes. DOE will provide public access to these results of federally sponsored research in accordance with the DOE Public Access Plan \href{http://energy.gov/downloads/doe-public-access-plan}. This work has been supported in part by DOE's Office of Electricity, in part by DOE's Building Technologies Office, and in part by NSF Award: ECCS-2145408.}
\thanks{$^{\dagger}$Xiang Huo was with the University of Utah, Salt Lake City, UT 84112 USA. He is now with the Department of Electrical and Computer Engineering, Texas A\&M University, College Station, TX 77843 USA (e-mail: {\tt\small xiang.huo@tamu.edu}).}%
\thanks{$^{\ddagger}$Boming Liu,
Jin Dong, and Jianming Lian are with the Electrification and Energy Infrastructures Division, Oak Ridge National Laboratory, Oak Ridge, TN 37830 USA
        (e-mail: {\tt\small liub,dongj,lianj@ornl.gov}).}%
        
\thanks{$^{\mathsection}$Mingxi Liu is with the Department of Electrical and Computer Engineering,
University of Utah,
Salt Lake City, UT 84112 USA
       (e-mail: {\tt\small mingxi.liu@utah.edu}).}%
       
}


\maketitle

\begin{abstract}
Reinforcement learning (RL)-based  methods have achieved significant success in managing grid-interactive efficient buildings (GEBs). However, RL does not carry intrinsic guarantees of constraint satisfaction, which may lead to severe safety consequences. Besides, in GEB control applications, most existing safe RL approaches rely only on the regularisation parameters in neural networks or penalty of rewards, which often encounter challenges with parameter tuning and lead to catastrophic constraint violations. To provide enforced safety guarantees in controlling GEBs, this paper designs a physics-inspired safe RL method whose decision-making is enhanced through safe interaction with the environment. Different energy resources in GEBs are optimally managed to minimize energy costs and maximize customer comfort. The proposed approach can achieve strict constraint guarantees based on prior knowledge of a set of developed hard steady-state rules. Simulations on the optimal management of GEBs, including heating, ventilation, and air conditioning (HVAC), solar photovoltaics, and energy storage systems, demonstrate the effectiveness of the proposed approach.


\end{abstract}

\begin{IEEEkeywords}
Distributed energy resources, grid-interactive efficient buildings, reinforcement learning, safe learning  
\end{IEEEkeywords}

\section{Introduction}

The optimal management  of grid-interactive efficient buildings (GEBs) is becoming more crucial than ever in reducing energy costs, integrating renewables, and facilitating grid decarbonization, owing to the fast-paced deployment of distributed energy resources (DERs). DERs, such as energy storage system (ESS), solar photovoltaic (PV), and electric vehicle (EV), together with flexible loads in GEBs, can be optimally controlled to provide revolutionary improvements in energy efficiency, grid resilience, and cost savings \cite{aguero2017modernizing,su2024renewable}. However, the growing modeling and computing complexity in GEB control, especially the explosion of  DERs, is posing unprecedented challenges for the safe and efficient management of GEBs.

To achieve the optimal management of GEBs, model-based methods that leverage scalable architectures have been thoroughly established \cite{jafari2022decentralized, pan2021distributed,huo2022two}. In \cite{jafari2022decentralized}, 
a decentralized control method was proposed to regulate the voltage in radial distribution networks by coordinating on-load tap changing  transformers and PV inverters.  Pan \emph{et al.} \cite{pan2021distributed} proposed a distributed low-communication algorithm for the control of islanded series of PV-ESS-hybrid systems. In 
\cite{huo2022two}, a two-facet scalable distributed algorithm was developed to offer scalability over both the agent population size and network dimension and verified through a residential EV charging control problem. In \cite{attarha2019affinely},  an affinely adjustable robust extension of the distributed alternating direction method of multipliers (ADMM) algorithm was designed to compensate for the deviations of forecasted loads and PV generation. Although model-based methods, including both distributed and decentralized control strategies, offer high scalability in GEB control problems, they can encounter impaired model accuracy and increased computing complexity, especially in complex building environments.

To deal with the growing system complexity, reinforcement learning (RL) has been prominently studied in power system applications owing to its model-free nature \cite{chen2022reinforcement}. RL-based control methods have the capability to comprehend hard-to-model  dynamics and surpass model-based methods in managing highly complex GEBs. In \cite{ye2020model}, a real-time model-free autonomous energy management strategy was proposed for residential multi-energy systems based on deep RL, where the user's energy cost is minimized under system uncertainties. In \cite{lu2022reward},  an actor–critic deep RL algorithm was developed based on reward shaping to manage the residential energy consumption profile with limited knowledge of the uncertain factors. In \cite{liu2020automated}, the control of the HVAC system is framed as a Markov decision process (MDP) using deep neural networks, and solved with deep deterministic policy gradient method to identify the optimal control strategy that reduces the energy cost and improves user comfort. 



The aforementioned RL-based approaches learn from trial-and-error interactions with the environment to maximize certain rewards. However, they generically neglect the safety considerations, i.e., ensure the safe behavior of RL agents in their environment. Consequently, the neglect of unsafe actions can lead to adverse consequences throughout both the online training and policy execution. Recently, safe learning methods are gaining more attention in industrial cyber-physical systems to emphasize the needs for real-world applications. A representative safe RL algorithm was proposed in \cite{bharadhwajconservative} to learn a conservative safety estimate of environment states through a critic, with the probability of failure in safety being upper-bounded. Admittedly, safe RL algorithms balance the tradeoff between safety and policy improvement, however, practical learing-based GEB control that offer customized or strict guarantees on certain hard constraints is still premature.

To fill this research gap, we aim at developing an RL-based GEB control algorithm with enhanced safety guarantees. The contributions of this paper are three-fold: 1) We develop an RL-aided GEB management framework to achieve optimized energy cost and consumer satisfaction; 2) The proposed physics-inspired safe RL minimizes room temperature constraint violations by projecting the actions onto a feasible region formed via a set of steady-state hard constraints; 3) The proposed method exhibits computational efficiency owing to its low computing cost in formulating steady-state feasible regions. Additionally, it is flexible in hard constraint selection and can be embedded in different RL structures. 




\section{System Model and Algorithm Design}
\subsection{System Model}
\label{System_Model}

\subsubsection{Dynamic building thermal network model}

The thermal dynamics are given based on the control-oriented resistance-capacitance (RC) thermal network model for a residential house with an attic \cite{cui2019load, huo2022twoGEB}. The thermal parameters are identified via a 4R4C thermal network model. 
The heat transfer within the building can be represented by the following system of first-order differential equations as
\begin{subequations}\label{6s}
\begin{align}
C_{in} \frac{dT^{in}_t}{d t}&=\frac{T^{w}_t-T^{in}_t}{R_{w}/2} + \frac{T^{a}_t -T^{in}_t}{R_{a}} + \frac{T^{m}_t-T^{in}_t}{R_{m}}  \nonumber\\ 
&+\frac{T^{amb}_t {-}T^{in}_t}{R_{win}} 
 {+} Q^{IHL}_t {-} c_1 Q^{AC}_t  {+} c_{2}Q^{sol}_t
\label{1}\\
C_{w} \frac{dT^{w}_t}{dt} &=\frac{T^{sol,w}_t-T^{w}_t}{R_{w}/2}-\frac{T^{w}_t-T^{in}_t}{R_{w}/2} 
\label{2}\\
C_{a} \frac{dT^{a}_t}{dt} &=\frac{T^{sol,f}_t-T^{a}_t}{R_{f}} {+}\frac{T^{a}_t-T^{in}_t}{R_{a}} {+}  c_4 T^{sol,a}_t\label{3} \\
C_{m} \frac{dT^{m}_t}{dt}&=-\frac{T^{m}_t-T^{in}_t}{R_{m}} + c_3Q^{sol}_t - c_5 Q^{AC}_t
\label{4} 
\end{align}
\end{subequations}where $t$ denotes the time index,  $C_{w}$, $C_{in}$, $C_{m}$, and $C_{a}$ denote the equivalent overall thermal capacitance of the exterior wall, indoor air, internal  mass, and air in attic, respectively. $R_{w}$, $R_{a}$, $R_m$, $R_{win}$, and $R_{f}$ denote the equivalent overall thermal resistance of exterior walls, attic floor, internal mass, window, and roof, respectively. $T^{in}_t$, $T^{w}_t$,  $T^{a}_t$, $T^{m}_t$, $T^{amb}_t$, $T^{sol,w}_t$, $T^{sol,f}_t$, and $T^{sol,a}_t$ are the indoor temperature, exterior wall temperature, attic air temperature, internal thermal mass temperature,  outdoor dry bulb temperature, and the effects of solar radiation on exterior walls, roofs, and attics, respectively. $c_{1}$, $c_{2}$, $c_{3}$, and $c_{4}$ denote the effective heating/cooling gain coefficients. $Q^{AC}_t$ denotes the cooling supply of the HVAC, $Q^{IHL}_t$ and $Q^{sol}_t$ denote the sensible heat gain from indoor heat sources and the solar radiation through windows, respectively.


 Assume each house is controlled by one HVAC, then the cooling supply of the $j$th HVAC from the $j$th house should stay within
\begin{equation}
    0 \leq Q^{AC}_{j,t} \leq \overline{Q}_{AC}, \forall j \in \Omega_{m}, \forall t \in \Omega_{t}
    \label{3ss}
\end{equation}
where $\overline{Q}_{AC}$ denote the cooling capacity limit, and $\Omega_{m}$ and $\Omega_{t}$ represent the set of houses and the set of time slots, respectively. 

To guarantee the indoor air temperature stays within the user's comfort range, $T^{in}_{t}$ should satisfy 
\begin{equation}
   \underline{T}^{in}
\leq   T^{in}_t \leq \overline{T}^{in}, \forall j \in \Omega_{m}, \forall t \in \Omega_{t}
   \label{5}
\end{equation}
where  $ \underline{T}^{in}$ and $\overline{T}^{in}$ denote the lower and upper temperature limits, respectively. 


\subsubsection{Solar photovoltaic} 

Solar PV can provide renewable power supply to an GEB by 
converting solar power into electricity, the active power injection $p^{s}_{j,t}$ from the PVs connected at house $j$ is limited by
\begin{equation}
0 \leq p^{s}_{j,t} \leq \overline{p}^{s}_{j,t}, \forall j \in \Omega_{m}, \forall t \in \Omega_{t}
\label{pv_limit}
\end{equation}
where $p^{s}_{j,t}$ denotes the active power injection, $\bar{p}^{s}_{j,t}$ denotes the maximum available active power from the associated PV inverter, and $\bar{p}^{s}_{j,t}$ is assumed to be known by the forecast. 

\subsubsection{Energy storage system} 

Assume the $j$th house is connected with an ESS that can offer the backup power supply through controlled charging and discharging actions. Let $p^{e}_{j,t}$ denote the controllable charging (positive) or discharging (negative) power that is in the range of 
\begin{equation}
\underline{p}^{dch}_{j} \leq p^{e}_{j,t} \leq \overline{p}^{ch}_{j}, \forall j \in \Omega_{m}, \forall t \in \Omega_{t}
\label{charge_limit}
\end{equation}
where $\underline{p}^{dch}_{j}<0$ and $\overline{p}^{ch}_{j}>0$ denote the maximum discharging and charging power limits, respectively. 

Besides, the capacity $E_{j,t}$ of the $j$th ESS subjects to the capacity limitation of 
\begin{equation}
\underline{E}_{j} \leq E_{j,t} \leq \overline{E}_{j}, \forall j \in \Omega_{m}, \forall t \in \Omega_{t}
\label{ESS_limit}
\end{equation}
where $\underline{E}_{j}$ denotes the allowable minimum energy stored in the ESS and $\overline{E}_{j}$
denotes the maximum energy capacity of the ESS.

\subsubsection{Objectives}

The objectives of the GEB control problem are formulated at optimizing the cost and comfort factors by 1) minimizing the total energy cost in the building; and 2) regulating the indoor room temperature to align closely with the predetermined setpoints. In specific, the energy cost minimization objective can be written as 
\begin{equation} \label{cost_obj}
  \mathcal{C}^{pr} = \sum_{j \in \Omega_{m}}  \sum_{t \in \Omega_{t}}  \left( p^{h}_{j,t} + p^{e}_{j,t} - p^{s}_{j,t}\right) \Delta t T^{pr}_{t}
\end{equation}
where $p^{h}_{j,t} = \delta Q_{j,t}^{AC}$ denotes the power consumption of the HVAC,  $\delta$ denotes the cooling coefficient of performance, $\Delta t$ denotes the time interval,  and $T^{pr}_{t}$ denotes the market electricity price. 

The customer comfort is measured by the indoor room temperature deviation as
\begin{equation} \label{tem_obj}
\mathcal{C}^{tem} = \sum_{j \in \Omega_{m}} \sum_{t \in \Omega_{t}}  \left(T^{in}_{j,t} - T^{set}_{j,t}\right)^{2}
\end{equation}
where $T^{set}_{j,t}$ denotes the predetermined temperature setpoint of house (customer) $j$.






\subsection{Algorithm Design}

\subsubsection{Markov decision process} 
\label{mdp}

To achieve the GEB optimization objectives while satisfying the system constraints, we formulate the energy building control problem as a Markov decision process (MDP) that aims at optimizing the energy cost and customer’s dissatisfaction. In specific, the MDP comprises four key components: a collection of states that represent the environment, a range of potential actions for each state, a reward function that evaluates the value of actions taken in specific states, and the transition probability among different states.

\textit{1) State:} The system state denoted by $s_t$ comprises system status that influence the decision-making process. In the GEB control problem, the system state at any time slot $t$ is defined as
\begin{align}
s_{t}=_{\times j \in \Omega_m}\left(T^{in}_{j,t},T^{set}_{j,t}, T^{pr}_{t}, T^{amb}_{j,t}, \overline{p}^{s}_{j,t}, E_{j,t} \right).
\end{align}
where $\times$ denotes the collection of all elements in a set. 

\textit{2) Actions:} Actions denoted by $a_{t}$ are operational decisions for controlling HVAC and DERs, and are defined by
\begin{align}a_{t}=_{\times j \in \Omega_m}\left(Q^{AC}_{j,t}, p^s_{j,t}, p^e_{j,t} \right).
\end{align}

\textit{3) Reward:} The reward consists of the negative sum of the electricity costs, the room temperature deviations, and the violation of the DER operational constraints 
\begin{align}\label{9s}
r_{t}=-\alpha_1 C_{t}^{pr} -  \alpha_2  C_{t}^{tem} -  \alpha_3  C_{t}^{s}
-  \alpha_4 C_{t}^{cd}  
-  \alpha_5 C_{t}^{ess}
\end{align}
where $\alpha_\varpi$, $\varpi=1,\ldots,5$, denotes the associated penalty coefficients.

The first term $C_{t}^{pr}$ calculates the energy cost by
\begin{equation}
  C_{t}^{pr} = \sum_{j \in \Omega_{m}} \left( p^{h}_{j,t} + p^{e}_{j,t} - p^{s}_{j,t}\right) \Delta t T^{pr}_{t}.
\end{equation}

The second term $C_{t}^{tem}$ measures the temperature deviation by
\begin{equation}
C_{t}^{tem} = \sum_{j \in \Omega_{m}}  \left(T^{in}_{j,t} - T^{set}_{j,t}\right)^{2}.
\end{equation}

The third term $C_{t}^{s}$ reflects the violation of solar PV limit via
\begin{equation}
C_{t}^{s} = \sum_{j \in \Omega_{m}}
 {\left(p^{s}_{j,t} - \bar{p}^{s}_{j,t} \right)}^2. 
\end{equation}

The fourth and fifth terms $C_{t}^{cd}$ and $C_{t}^{ess}$ assess violations on the ESS charging/discharging power limits and ESS capacity limit, respectively, and they are calculated by 
\begin{align}
  C_{t}^{cd} &{=} \sum_{j \in \Omega_{m}} \max (0, p^e_{j, t} -\overline{p}_{j}^{ch}) {+} \max (0, \underline{p}_{j}^{dch} - p^e_{j, t}) \nonumber\\
  C_{t}^{ess} &{=} \sum_{j \in \Omega_{m}} \max (0, E_{j, t} -\overline{E}_{j}) {+} \max (0, \underline{E}_{j} - E_{j,t}). 
\end{align}

In an attempt to optimize cost-effectiveness, ESSs will strategically charge when market electricity prices are low, while inversely, discharge is preferred during periods of high  electricity prices. The solar PVs are stimulated through rewards to inject the solar energy into the grid to meet the load demand. Any superfluous solar energy generation will be stored in the ESS for future utilization. The metric of customer dissatisfaction is quantitatively represented as the discrepancy between the actual room temperature and the scheduled temperature setpoints.

\subsubsection{Steady-state analysis}

Inspired by the physics of HVAC system, we propose a safe RL-based learning method that incorporates physics principles to achieve enhanced safety guarantees. The heat transfer within the building is first formulated into a steady-state problem, aiming at tracking desired temperature setpoints. Then we develop a set of indoor room temperature constraints based on the steady-state analysis to regulate the constraint violations. The cooling supply from HVAC will be controlled to drive the system states to a series of steady states that are the optimal solutions. For simplicity, we only consider the summer cooling scenario when the HVACs are always on.

\noindent\textbf{Proposition 1}. When the HVAC system is on, the linear system in \eqref{6s} converges asymptotically, and the equilibrium state is uniquely determined by the input $\mathbb{T}$ $\triangleq$ $\{Q^{AC}_t, T^{amb}_t, Q^{IHL}_t, Q^{sol}_t, T^{sol,w}_t, T^{sol,f}_t, T^{sol,a}_t\}$.  \hfill $\square$

Proposition 1 can be verified by representing the linear system in \eqref{6s} into 
a continuous state space model as 
\begin{subequations}\label{5ss}
\begin{align}
    \dot{\bm{\mathcal{X}}}_{\tilde{t}} &= \bm{A}\bm{\mathcal{X}}_{\tilde{t}} + \bm{B}Q^{AC}_{\tilde{t}} + \bm{G}\bm{\mathcal{D}}_{\tilde{t}}\label{14a}\\
    \bm{\mathcal{Y}}_{\tilde{t}} &= \bm{C}\bm{\mathcal{X}}_{\tilde{t}} \label{14b}
\end{align}
\end{subequations}
where
\begin{align}
\bm{A}&=
\begin{bmatrix}\frac{-2}{R_w} {-} \frac{1}{R_a} {-}   \frac{1}{R_m} {-} \frac{1}{R_{win}} & \frac{2}{R_w} &\frac{1}{R_a} &\frac{1}{R_m}\\
\frac{2}{R_w} & \frac{-4}{R_w} & 0 & 0 \\
\frac{-1}{R_a} & 0 & \frac{1}{R_a} - \frac{1}{R_f} & 0\\
\frac{1}{R_m} & 0 & 0 & \frac{-1}{R_m}
\end{bmatrix} \nonumber\\
\bm{B}&=
\begin{bmatrix}-c_1\\
0 \\
0  \\
-c_5\\
\end{bmatrix}, \bm{G}=
\begin{bmatrix}\frac{1}{R_{win}} &1 &c_2 & 0 & 0 & 0\\
 0 & 0 & 0 & \frac{2}{R_w} & 0 & 0 \\
 0 & 0 & 0 & 0  & \frac{1}{R_f} & c_4 \\
 0 & 0 & c_3 & 0  & 0 & 0 \\
\end{bmatrix},\nonumber
\end{align}
$\bm{\mathcal{X}}_{\tilde{t}} = [T^{in}_{\tilde{t}}, T^{w}_{\tilde{t}},  T^{a}_{\tilde{t}}, T^{m}_{\tilde{t}}]^{\mathsf{T}}$ denotes the system state, $\bm{\mathcal{Y}}_{\tilde{t}}$ denotes the system output, $\bm{\mathcal{D}}
_{\tilde{t}} = [T^{amb}_{\tilde{t}}, Q^{IHL}_{\tilde{t}}, Q^{sol}_{\tilde{t}}, T^{sol,w}_{\tilde{t}}, T^{sol,f}_{\tilde{t}}, T^{sol,a}_{\tilde{t}} ]^{\mathsf{T}}$ denotes the constant environment input, $\bm{C} \in \mathbb{R}^{4\times4}$
denotes an identity matrix, and $\tilde{t}$ denotes the steady-state time index for clarity. The controllability of system \eqref{5ss} can be directly verified using Kalman’s test.

Therefore, when the HVAC system is on, the objective aims at designing the dynamics of $Q^{AC}_{\tilde{t}}$ to achieve desired steady states that can track the predefined temperature points \cite{zhang2017decentralized}. The HVAC supply is controlled as input to maintain the indoor air temperature as close as possible to the setpoint. At any steady state, \eqref{14a} becomes
\begin{equation}
\bm{A}\bm{\mathcal{X}}_{\tilde{t}} + \bm{B}Q^{AC}_{\tilde{t}} + \bm{G}\bm{\mathcal{D}}_{\tilde{t}}= \bm{0}.
    \label{7}
\end{equation}
Therefore, we can readily obtain that  
\begin{equation}
\bm{\mathcal{X}}_{\tilde{t}} = - \bm{A}^{-1}( \bm{B}Q^{AC}_{\tilde{t}} + \bm{G}\bm{\mathcal{D}}_{\tilde{t}}).
    \label{7s}
\end{equation}
Consequently, the indoor temperature equals to  
\begin{equation}
T_{\tilde{t}}^{in} = \frac{c_1+c_2}{\tau} Q^{AC}_{\tilde{t}} + \frac{\omega}{\tau} 
\label{8}
\end{equation}
where $\tau = 1/(z_2r_a^2) - 1/r_w - 1/r_a - 1/r_{win}$, $\omega_{\hat{\imath}} = \text{row}_{\hat{\imath}}(G)\mathcal{D}_{\tilde{t}}$, $\forall \hat{\imath}=1,\ldots,4$, $\text{row}_{\hat{\imath}}(\cdot)$ denotes the $\hat{\imath}$th row of a matrix, and $\omega = \omega_1 + \omega_2/2 - \omega_3/(r_a z_2) + \omega_4$. 

Without loss of generality, we take the room temperature constraint for example to show the regulation on GEB temperature control. The cooling supply of an HVAC system at time $\tilde{t}$ should satisfy 
\begin{equation}
\Psi_{\tilde{t}} \triangleq \{ Q^{AC}_{\tilde{t}} \mid \eqref{8}, \eqref{3ss}, \eqref{5} \}
\end{equation}
where $\Psi_{\tilde{t}}$ denotes the feasible region of $Q^{AC}_{\tilde{t}}$ at time $\tilde{t}$.

By substituting the cooling capacity limit in \eqref{3ss} and the indoor temperature constraint in \eqref{5}, the feasible region of the cooling supply $\Psi_{\tilde{t}}$ can be explicitly written into 
\begin{equation} \label{21s}
\max \{0, \frac{\tau(\omega+\overline{T}^{in})}{c_1+c_2} \} {\leq} Q^{AC}_{\tilde{t}} {\leq} \min \{\frac{\tau(\omega+\underline{T}^{in})}{c_1+c_2},  \overline{Q}^{AC} \}.
\end{equation}

Compared with the original cooling supply constraint in \eqref{3ss}, the feasible region in \eqref{21s} confines the operation of an HVAC to achieve optimal steady states. With the assistance of physical knowledge about the HVAC and the control requirements, the feasible region can be self-defined to regulate the indoor room temperature to the optimal or desired setpoints.

\subsubsection{Safe layer design}

In GEB control problems, RL agents often operate in dynamic and uncertain building environments where unsafe actions can lead to detrimental consequences. In this section, an external safety layer is integrated into the RL design to verify the safe actions of HVACs before each execution, ensuring temperature constraint guarantees. We prioritize the safe temperature constraints by using the developed steady-state feasible region to achieve enhanced constraint satisfaction when controlling GEBs. 

\begin{figure}[thb]
    \centering
	\includegraphics[trim={0cm 0cm 0cm 0cm}, clip,width=0.95\linewidth]{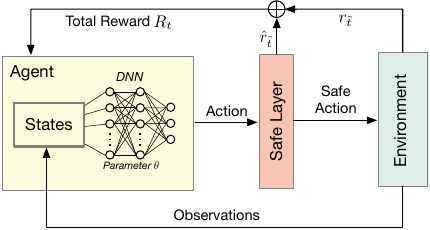}
	\caption{Physics-inspired safe RL structure.}
	\label{fig_safe_RL}       
\end{figure}

In specific, the derived steady-state constraints of the cooling supply contour a feasible region, i.e., $\Psi_{\tilde{t}+1}$, of the next action. When the HVAC action stays within  $\Psi_{\tilde{t}+1}$, it is considered a safe action. If any action that poses a risk falls outside the feasible region, it will be redirected to the action within the region that has the shortest Euclidean distance. The architecture of the proposed safe RL algorithm is shown in Fig. \ref{fig_safe_RL}. 

Note that the transition to the next state is not observed under the unsafe action, as such an action  is not actually executed. However, exploration in such an unfeasible area is allowed. To penalize the potential constraint violations by unsafe actions, the following safe-layer penalty is added as 
\begin{equation}
\hat{r}_{\tilde{t}} = - \hat{\alpha} \|\hat{a}_{\tilde{t}} - a_{\tilde{t}} \|_2
\end{equation}
where $\hat{a}_{\tilde{t}}$ denotes the safe action in $\Psi_{\tilde{t}+1}$ that has the shortest Euclidean distance to $a_{\tilde{t}}$, i.e., $\min_{\hat{a} \in \Psi_{\tilde{t}+1}} \|\hat{a} - a_{\tilde{t}} \|_2$, and $\hat{\alpha}$ denotes the penalty coefficient. Note that $\hat{r}_{\tilde{t}}$ expects to train the RL algorithm to prioritize selecting the safe actions for the HVAC, therefore can lead to lower rewards for $r_{\tilde{t}}$. 

Therefore, the total reward after adding the safe-layer penalty becomes 
\begin{equation}
R_{\tilde{t}} = r_{\tilde{t}} + \hat{r}_{\tilde{t}}.
\end{equation}

The detailed procedures of the proposed safe RL algorithm are shown in Algorithm \ref{alg_1}.


\noindent \textbf{Remark 1}. The safe layer design of Algorithm \ref{alg_1} is inspired by exploring the thermal dynamics of GEBs in steady states.  To apply  Algorithm \ref{alg_1}, we assume that the hard constraints are prior knowledge that aids in refining the policy and providing safety guarantees. In this paper, the equivalent thermal model RC parameters are obtained through building model identification, and used directly in the environment to demonstrate the performance of safe actions. However, the feasible region $\Psi_{\tilde{t}}$ also can be formulated separately from the RL algorithm based on practical control requirements on the actions. \hfill $\blacksquare$

\RestyleAlgo{ruled} 

\begin{algorithm}
\caption{Physics-inspired GEB control with enhanced safe RL}
\label{alg_1}
\textbf{Initialization:} Initialize the environment, replay memory $\mathbb{R}$, Q-network, and target Q-network with random weights $\theta$ and $\hat{\theta}$, exploration rate $\epsilon$; Initialize the states: $s_{0}= \{T^{in}_{j,0}, T^{set}_{j,0}, T^{pr}_{0}, T^{amb}_{j,0}, \overline{p}^{s}_{j,0}, E_{j,0} \}$;

\For {$episode = 1, M$}{
Reset the environment $s_{\tilde{t}}=s_{0}$;

\For {${\tilde{t}} = 1 , T$}{With probability $\epsilon$, select a random action $a_{\tilde{t}}$, 

otherwise, select $a_{\tilde{t}} = \arg\max_{a}Q(s_{\tilde{t}},a;\theta)$;
 
 \eIf {$a_{\tilde{t}} \not\in \Psi_{\tilde{t}}$ }{
  $\hat{a}_{\tilde{t}}=\Pi_{\Psi_{\tilde{t}}}(a_{\tilde{t}})$, set $\hat{r}_{\tilde{t}} = - \hat{\alpha}\|\hat{a}_{\tilde{t}} -a_{\tilde{t}} \|_2$;}
  {keep $\hat{a}_{\tilde{t}} = a_{\tilde{t}}$ as the safe action, set $\hat{r}_{\tilde{t}} = 0$;}

  Execute $\hat{a}_{\tilde{t}}$ on HVAC and obtain the reward $r(s_{\tilde{t}}, {\hat{a}}_{\tilde{t}})$ and next state $s_{\tilde{t}+1}$ ;
  
  Store the transition $(s_{\tilde{t}}, a_{\tilde{t}}, r_{\tilde{t}}, s_{\tilde{t}+1}, \text{done})$ in $\mathbb{R}$;
   
  Sample a mini-batch transitions $(s_i, a_i, r_i, s_{i+1}, \text{done})$ from $\mathbb{R}$ ;
  
  Set $y_i =\begin{cases} 
  r_{i} & \text{done} {=} 1,\\
r_{i} {+} \gamma \max_{a_{i+1}}\hat{Q}(s_{i+1},a_{i+1};\hat{\theta}) &  \text{done} {=} 0;
    \end{cases}$

Perform gradient descent on $(y_{i}{-}Q(s_{i},a_{i};\theta))^{2}$;

Decay $\epsilon$;

Set $\hat{Q} = Q$ every $k$ steps;
  }
 }

\end{algorithm}

\subsubsection{Deep Q-networks} 

Q-learning is a model-free RL technique that aims to learn an optimal action-value function, also known as the Q-function, which estimates the expected cumulative reward for taking a particular action in a given state. Deep Q-networks (DQNs) combines deep neural networks with Q-learning by using deep neural networks as function approximators to approximate the Q-function \cite{mnih2013playing}.  

The state of the system at each time step is represented as inputs to the neural network. In this paper, the state includes room temperatures, outdoor temperatures, electricity price, time of the day, maximum available active power from PVs and the energy statues of the ESSs. The neural network consists of multiple fully connected layers of neurons to take the states as inputs and outputs the Q-values for each possible action. The output layer has one neuron for each possible action, representing the Q-value estimate for each action. During training, it explores the environment by selecting random actions with an exploration rate $\epsilon$. As training progresses, $\epsilon$ decreases and the DQN tends to exploit its learned knowledge by selecting actions with the highest Q-values.

\begin{figure}[!thb]
\vspace{-2mm}
    \centering
\includegraphics[width=0.48\textwidth]{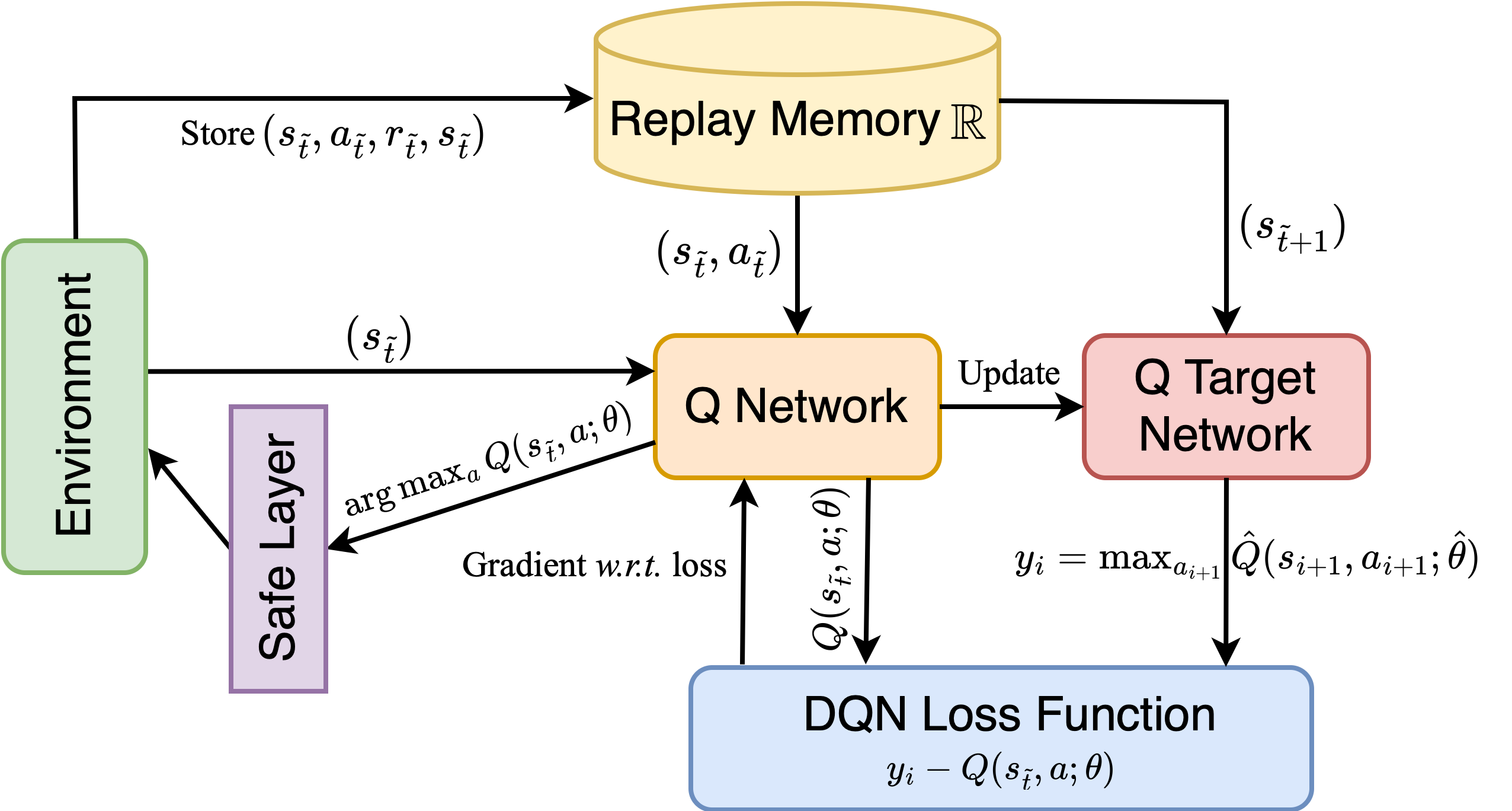}
	\caption{The structure of the proposed safe-RL network based on DQN.}
	\label{fig_structure_network}       
 \vspace{-2mm}
\end{figure}

To improve the sample efficiency and reduce correlations between consecutive samples, experience replay is employed. Mini-batches of samples, i.e., $(s_i, a_{i}, r_i, s_{i+1}, \text{done})$, are randomly sampled from the replay memory $\mathbb{R}$ to train the neural network. By iteratively updating the weights $\theta$ of the neural network based on the training, the model learns to approximate the optimal Q-function and make informed decisions about the HVAC cooling supply, solar PV power injection, and ESS charging/discharging power. The structure of the proposed safe-RL network using DQN is shown in Fig. \ref{fig_structure_network}. 

Algorithm \ref{alg_1} adopts a DQN as the function approximator to calculate the Q-values, the Q-function gives the expected return of taking a particular action $a$ in a given state $s_{\tilde{t}}$. A discount factor $\gamma$ is used to balance the importance of immediate rewards versus future rewards in the decision-making process. With an experienced memory, transitions are stored and sampled randomly for training. During each episode, the agent selects actions based on an $\epsilon$-greedy exploration strategy to balance exploration and exploitation. The Q-network is updated by minimizing the mean squared error loss between the predicted Q-values and the target Q-values, which is $(y_{i}-Q(s_{i},a_{i};\theta))^{2}$. The target network $\hat{Q}$ is updated every $k$ steps by copying the weights from the online network $Q$ to enhance learning stability and improve the convergence of the algorithm. The algorithm iteratively repeats episodes until convergence. 

\section{Simulations}

In this section, we evaluate the performance of the proposed safe-RL method on an GEB control problem. The objective is to minimize the total energy cost and maximize user comfort as described in \eqref{cost_obj} and \eqref{tem_obj}. In the simulation setting, one ESS is assumed to be connected at each house with a capacity of $2$ kWh and a maximum charging/discharging power of $\pm1$ kW; Additionally, it is presumed that the ESS has a charging and discharging efficiency of $0.98$ and $0.85$, respectively, and should preserve allowable minimum energy of $0.3$ kWh as backup power for emergency. Solar PV generation was estimated on a sunny day and scaled to have a peak power of $0.3$ kW \cite{NREL_solar}. 

The time interval is set to be $\Delta t= 15$ mins on a daily basis.  The day-ahead electricity price is estimated based on data from the Atlanta electricity market \cite{Atlantic_electricity}. We adopt the 4R4C model whose RC parameters are identified to be $C_{w} = 10,000,000$, $C_{in} = 329,472$, $C_{m} = 14,644,976$, $C_{a} = 2,330,670$, $R_{w} =0.0057$, $R_{a} = 0.2$, $R_m = 0.1$, $R_{win} =0.0807$, and $R_{f} = 0.0965$, respectively.  The effective heating/cooling gain coefficients are
$c_{1} = 0.5$, $c_{2} = 0.5$, $c_{3} = 0.4$, $c_{4} = 0.8$, and $c_{5} = 0.5$, respectively.  For the training of DQN, the learning rate $\alpha$ is set to be $0.001$. The exploration rate is initially set as $\epsilon = 1$ and then linearly decays over time. The discount factor is set as $\gamma = 0.99$. The replay memory size is $10,000$, and the sample batch size for each training iteration is $64$. The Q-networks has two hidden layers with $256$ hidden units for each layer.   
\begin{figure*}
\setkeys{Gin}{width=1\linewidth}
\begin{minipage}[t]{0.24\textwidth}
\includegraphics{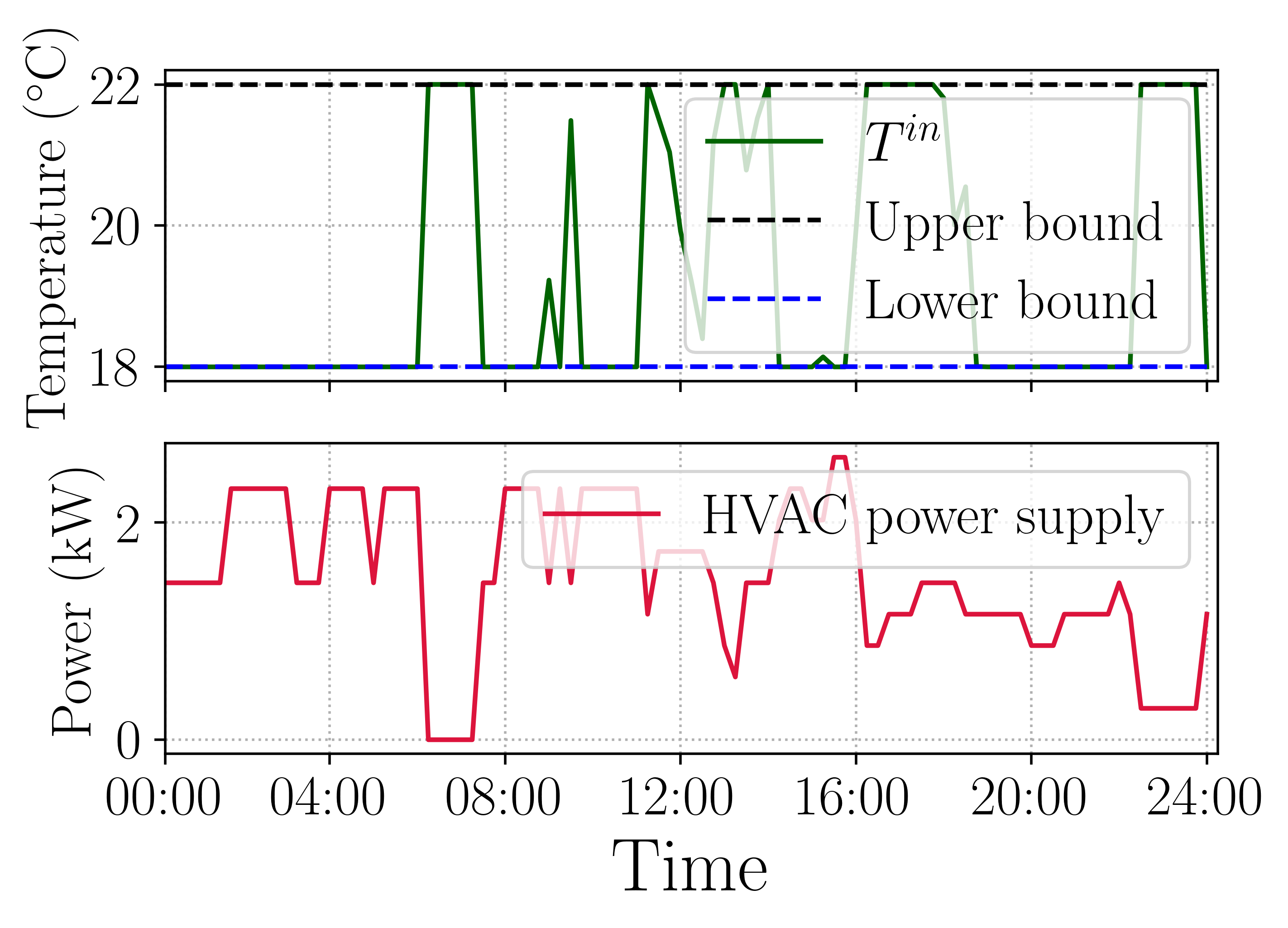}
\caption{Indoor room temperature and HVAC power supply.}
\label{fig_HVAC_actions}
\end{minipage}\hfill
\begin{minipage}[t]{0.24\textwidth}
\includegraphics{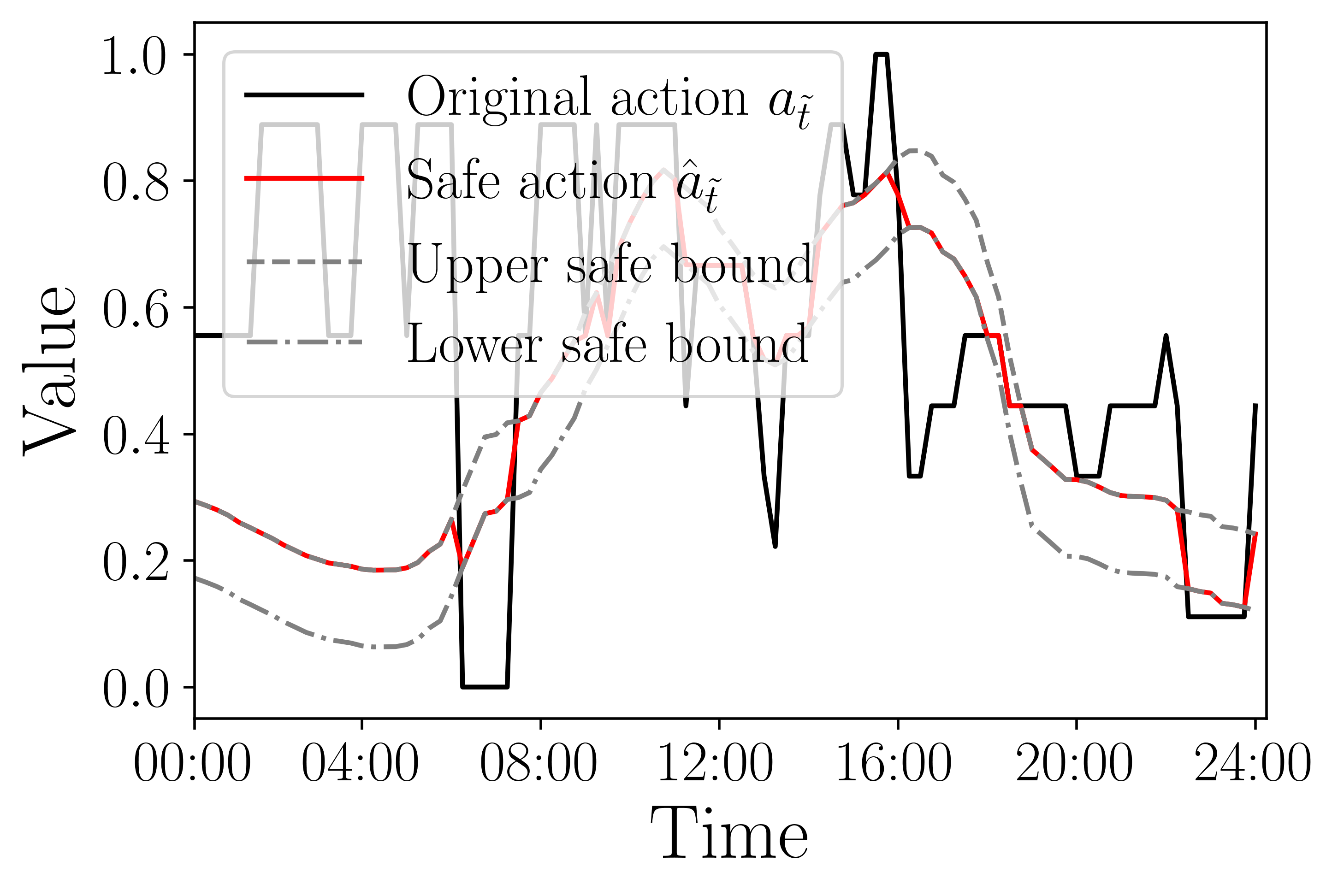}
\caption{Original actions and regulated safe actions during the HVAC control.}
\label{fig_safe_and_unsafe_actions}
\end{minipage}\hfill
\begin{minipage}[t]{0.24\textwidth}
\includegraphics{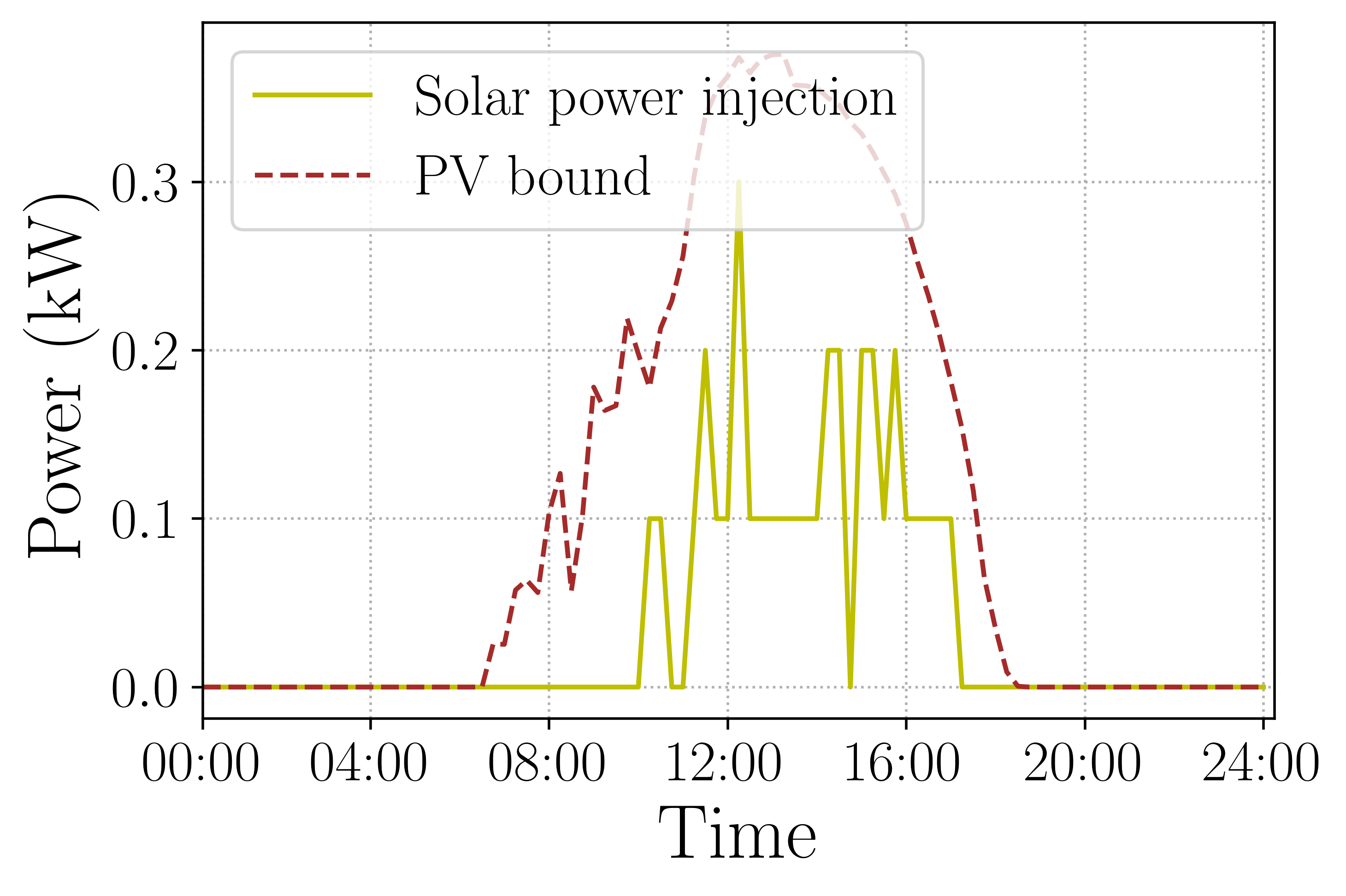}
\caption{Solar power injection from the solar PVs.}
\label{fig_pv} 
\end{minipage}\hfill
\begin{minipage}[t]{0.24\textwidth}
\includegraphics{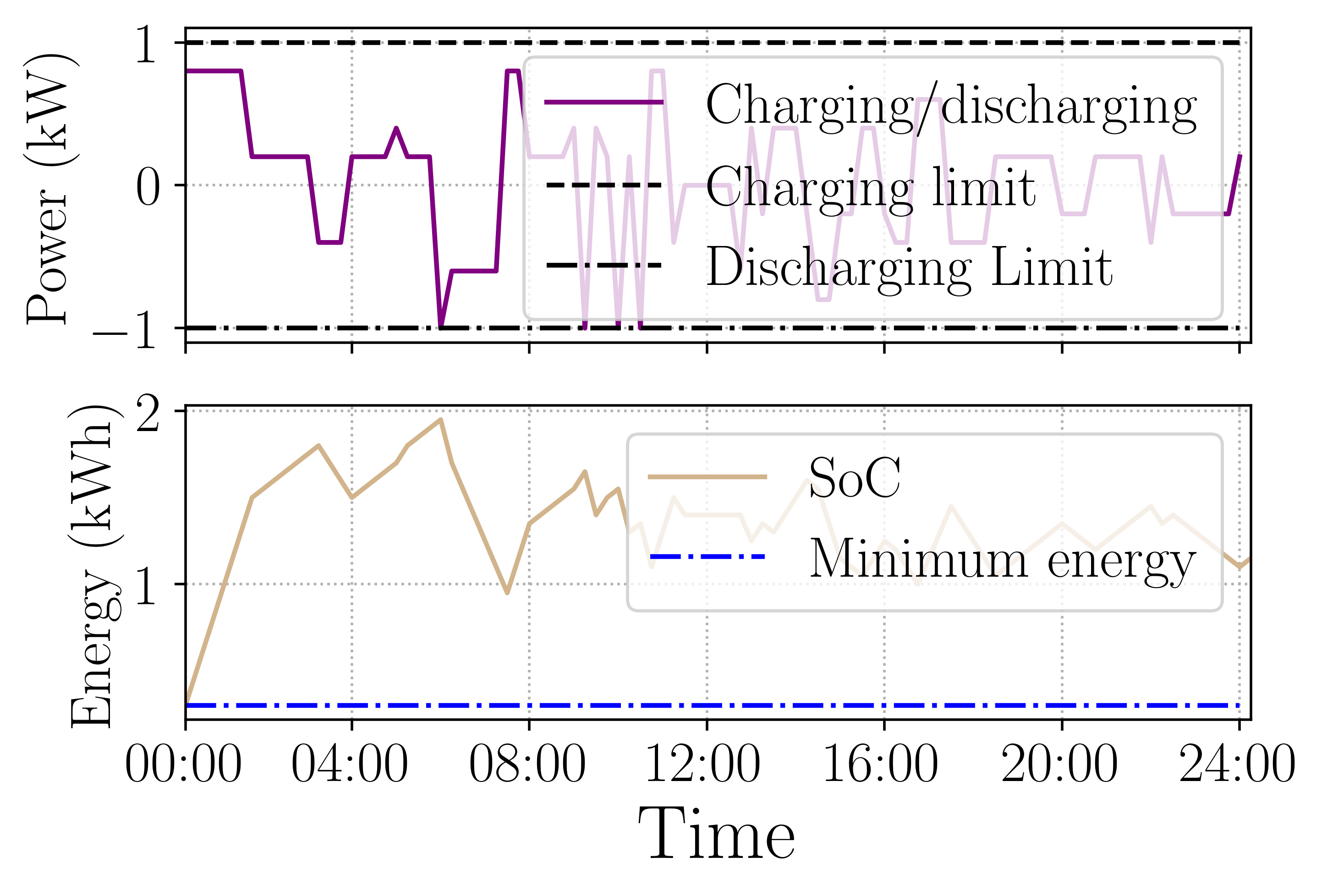}
\caption{Charging/discharging schedules of the ESS.}
\label{fig_ESS} 
\end{minipage}
\vspace{-4mm}
\end{figure*}
Fig. \ref{fig_HVAC_actions} 
demonstrates the indoor air temperature control results with proposed safe RL algorithm. While the HVAC is on cooling status throughout the simulation, the indoor room temperature is strictly regulated within $[18,22]$ $^{\circ}$C by using the feasible region developed in steady states. Note that the precise temperature control is only obtained as a result of the exact RC model parameters. As shown in Fig. \ref{fig_safe_and_unsafe_actions}, 
the original actions before entering the safety layer violate the safety bounds drastically. After applying the proposed safe RL, unsafe actions were regulated to stay within the feasible region, leading to enhanced temperature bounds guarantees.  

\begin{figure}[thb]
\vspace{-2mm}
    \centering
	\includegraphics[trim={0cm 0cm 0cm 0cm}, clip,width=0.96\linewidth]{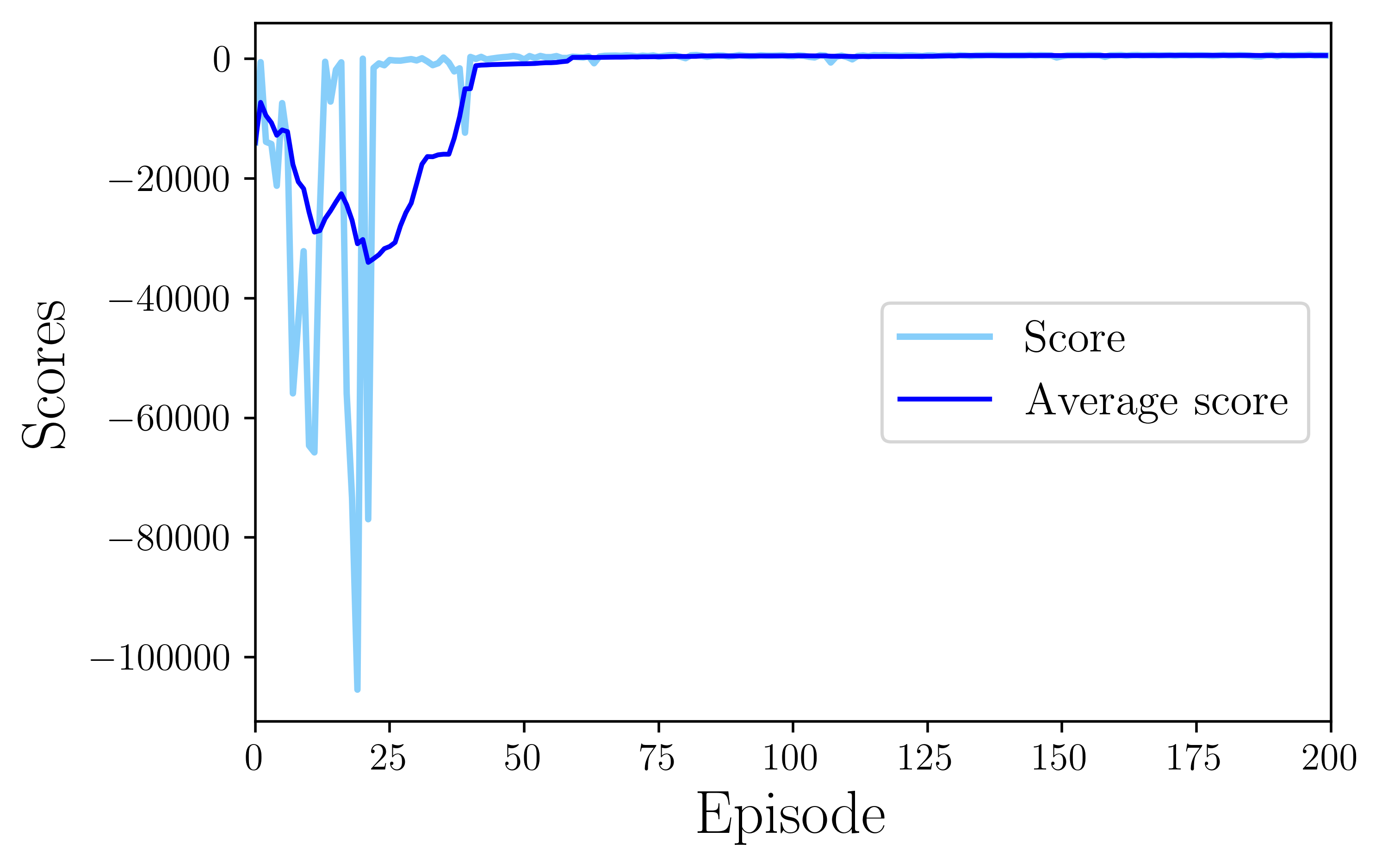}
  \vspace{-2mm}
	\caption{Score and average scores in the RL process.}
	\label{fig_scores}       
 \vspace{-3mm}
\end{figure}

In Fig. \ref{fig_pv}, 
the solar power has a peak power generation around noon, and was maximumly utilized in building electricity supply to help reduce the energy cost. Fig. \ref{fig_ESS} 
reflects the charging/discharging behaviors of the ESS. The ESS tends to store more energy off the peak hour, e.g., before 8 A.M., and discharge to provide electricity whenever needed. Besides, the capacity of the ESS is always above the lower capacity limit $0.3$ kWh. Finally, Fig. \ref{fig_scores}
 measures the performance of the agent's interactions with the environment. The agent learns the optimal GEB control policy that maximizes the cumulative reward over time, including the cumulative score during a specific episode and average scores from all episodes.


\section{Conclusion}

This paper presents a physics-inspired safe-RL framework for the optimal management of DERs and HVACs in GEBs. Different energy-consuming and producing resources are coordinated to optimize energy usage, reduce electricity costs, and improve customer comfort. The proposed safe-RL approach can achieve an enhanced safety constraint guarantee based on  the physics of the HVAC system for indoor room temperature control. The developed method also eliminates the computational power overhead compared to solving a dynamic optimization problem. The simulation studies on the GEB problem show the effectiveness of the proposed method in learning to control diverse energy resources in a safe and cost-efficient manner. Future work includes contemplating the trade-off between reward and safety, when applied to large-scale cyber-physical energy systems.

\bibliographystyle{IEEEtran}

\bibliography{bibliography}

\end{document}